# Editorial: Hierarchical colloidal nanostructures – from fundamentals to applications

Christian Klinke* (Guest editor)

Institute of Physical Chemistry, University of Hamburg, Martin-Luther-King-Platz 6, 20146 Hamburg, Germany

In order for our society to stay competitive in terms of computing power, sensing, and alternative energy new materials are required. Material science is one of the pillars of innovation in modern culture. In respect of future optical and electrical applications there is a need for active components with higher efficiencies and lower costs. Nanostructured materials such as nanoparticles, nanowires, carbon nanotubes, graphene, and nanosheets are considered to be among the most promising candidates for faster and less expensive electronic devices and more efficient solar cells and fuel cells. This is mostly due to their low-dimensional nature which opens new intriguing possibilities for technology. The synthesis of inorganic nanostructures defined by colloidal chemistry has developed strongly and became a versatile tool in various fields of application: Colloidal nanoparticles are used as fluorescence markers in medicine and biology and in the latest generation of TV sets semiconducting nanoparticles are used in the background illumination systems, which demonstrates the bright future of colloidal nanomaterials. Further, colloidal nanostructures of various dimensionality and hierarchy can be used as inexpensive transistors, photo-detectors, chemical sensors, thermoelectric materials, and as solar cells. Exciting properties include Coulomb blockade, ballistic transport over micrometers in length, and continuously tunable optical properties. In order to produce nanomaterials with new, tailored properties it is indispensable to understand the mechanisms of synthesis, optical processes and electrical transport in detail.

In this issue of "Zeitschrift für Physikalische Chemie" a representative cross-section of researchers active in Germany or with close relations report on their interesting, recent results in synthesis, assembly, and characterization. In a first contribution, the group of Victor Puntes demonstrates a simple room-temperature approach for the synthesis of gold nanoparticles and how to gain control over size in this one-pot synthesis [1]. They discuss a nice model for the steps of reduction of the gold ions by the two used reducing agents, the nucleation of seeds and the growth of the nanoparticles. The degree of control of the number of seeds and the size of nanoparticles in this kind of synthesis is unprecedented. Gold nanoparticles are interesting due to their optical properties arising for a plasmonic resonance which is a function of the size of the nanoparticles. In this respect, a coupling of the nanoparticles detunes the resonance and spectroscopy can be correlated with the formation of super-structures. In their study, Holger Lange and co-workers investigate the agglomeration of gold nanoparticles in different solvents and using different linker molecules to heap-like or chain-like structures by comparison of UV-vis spectroscopy, dynamics light scattering, and transmission electron microscopy [2]. The knowledge about the aggregation of nanoparticles can guide the synthesis of defined, complex structures which could be exploited for plasmon-based sensing. Beyond aggregation, also dissolution, oxidation, and adsorption of molecules on the nanoparticles leave traces in the absorption spectra. Neus Bastus et al. show in their paper that a wealth of properties can be extracted from such measurements [3], despite the simplicity of the method in terms of preparation and measurement affort. Additionally, spectroscopy can be applied in-situ, e.g. in flow reactors. The parameter that can be extracted form UV-vis absorption spectra are: changes in the extinction intensity (of the plasmon resonance), a shift in the plasmon peak position, the full-width half-maximum value, and the appearance of additional features. By modeling the optical properties of nanoparticle solutions, the authors carefully analyze the correlation of these parameters with the processes of nucleation, growth, aggregation of the particles and discuss the limits of this method (e.g. parallel processes).

The applicability of plasmonic nanoparticles is not only limited to spectroscopic properties. The group around Nikolai Gaponik and Alexander Eychmüller shows that gold nanoparticles can serve as catalyst for the CO oxidation [4]. For that, they decorate titanium oxide particles with gold nanoparticles. Usually, the stabilizing ligand shell around the nanoparticles hampers the catalytic processes. In order to circumvent these difficulties, before casting the nanoparticles on the larger $TiO_2$ support particles they perform a ligand exchange from the original citrate molecules to tetrazoles. The advantage of latter is that, after immobilization of the gold particles on the support, they decompose to gaseous fragments upon heating, leaving the gold particles virtually ligand-free. After the heating step, the materials show a much higher catalytic activity, a proof for thermal catalyst activation under mild conditions under preservation of the nanoparticles' catalytically active size range of several nanometers. The catalytic activity is also guaranteed by the hybrid, hierarchical structure of the combination of larger $TiO_2$ particles with smaller gold nanoparticles, leaving voids for gaseous mass transport. The concept of open, hierarchical structures for heterogeneous catalysts further developed by the group of Nadja Bigall [5]. They produce aerogels by cryogelating water-soluble metal nanoparticles followed by freeze drying. This procedure yields highly porous agglomerates which combine structural stability with an easy access for gaseous reactants. This is demonstrated for the oxidation of carbon monoxide on various catalytic nanoparticle systems (Pd, Pt, Ag, Au). Beyond, these catalysts show better performance comparted to traditional supported catalysts in terms of heat transfer and active material per total weight including the support, though the thermal stability is limited (with the exception of Pd). Anyhow, the presented results demonstrate the potential of hierarchical nanostructures for future applications, especially in the billion Euro market for catalysts. Since catalysts are relevant in a large variety of industrial processes and their application can save enormous amounts of energy, their further development is vital for the prosperity of our society.

To turn to optically interesting, semiconducting nanostructures, Klaus Boldt reviews the process of improvement of their spectroscopic properties, in particular in terms of fluorescence [6]. The most successful method to do so is to grow a shell of another semiconductor material on the nanoparticles which stabilizes the properties (type I). Anyhow, at the interface between the two materials of the core/shell particles there is an abrupt, theta-function-like change in the band structure leading to a sub-optimal optical performance. As a remedy, he introduces the possibility of a gradual change in the material composition. The influence of the synthesis conditions is discussed, as well as additional annealing steps, the electronic transition mechanisms, and the possibility of deliberate band engineering. Latter is also the notion in the following article by Dirk Dorfs and his group, who introduce UV-vis spectroscopy as a slick tool to determine the geometrical properties of CdSe/CdS core-shell nanorods (interestingly similar to the intention of Neus Bastus in the case of metal nanoparticles) [7]. In these hybrid structures the features of core and shell are located at different wavelengths in the spectra and thus, can be analyzed separately. In this way, using careful calibrations the CdSe core diameter can be determined from the lowest energy absorption maximum, the rod width can be determined by the position of the CdS related absorption band, the overall rod volume can be estimated by the ratio of the maximum extinction values of the two aforementioned absorption bands, and the particle concentration can be determined by the optical density at high energy. Due to their superior properties, such core-shell and graded-shell particles are considered to be integrated into passive or even active illumination systems in future flat panel display.

An interesting young development in the synthesis of colloidal nanomaterials is the possibility of two-dimensional structures. Usually, such nanosheets are stabilized by protecting ligands which self-assemble to a monolayer around the inorganic 2D part [8]. The confinement is present only in one direction, namely in the height nanosheets. This opens the possibility of charge carrier movement in the plane of the colloids, while still having the possibility to tune the bandgap by the confinement in height. Latter allows an optimization of the properties for relevant applications like solar cells [9]. Recently, Jannika Lauth succeeded in synthesizing InSe nanosheets which are less toxic compared to the working horse materials CdSe [10]. In her contribution with the group of Laurens Siebbeles, she investigates the optoelectronic properties, in particular the contact-less mobility, of these InSe nanosheets [11]. By means of transient absorption spectroscopy the recombination kinetics was studied and THz spectroscopy showed that the mobility of free charge carriers can reach respectable 20 $cm^2/Vs$, keeping in mind that the nanosheets possess a height of the inorganic part of only about 1 nm. Eventually, colloidal nanostructures should be applied as inexpensive, high quality optoelectronic materials. In regards to solar cells, Alina Chanaewa and Elisabeth von Hauff studied the electrical transport and the dielectric loss in PbS nanoparticle films by means of impedance spectroscopy [12]. They found that trap states have a decisive impact on the electronic properties and that ligand exchange treatment can remedy the problem to some extent. Based on the results the mechanisms of trap formation and trap filling are discussed which is very helpful for the further development of useful electronic devices based on colloidal nanoparticles. Even more forward-looking is the approach of Marcus Scheele and his group who functionalized the PbS nanoparticles with photosensitive organic ligands and thus, prepared an strongly coupled nanoparticle film [13]. Upon illumination with appropriate wavelength, the chosen ligand can open and close its conformation in solution. In solid state this is much more difficult and after switching usually one configuration is dominant. Anyhow, the authors find remarkable photoconductivity and fast photoresponse in those films. Further, they discuss the next necessary steps to improve such devices to be useful as photoswitchable transistors or memory devices.

In summary, this issue of "Zeitschrift für Physikalische Chemie" assembles articles from various younger groups active in the field of synthesis of colloidal nanostructures, their fundamental characterization, and applied concepts for future devices. Enjoy reading the presented contributions. They demonstrate strong, trendsetting activities in this branch of material science and might contribute to the nowadays societal challenges.

## References

1. R. A. Sperling, L. Garcia-Fernandez, I. Ojea-Jimenez, J. Piella, N. G. Bastus, and V. Puntes, Z. Phys. Chem. 231 (2017) 7.
2. M. Deffner, F. Schulz, and H. Lange, Z. Phys. Chem. 231 (2017) 19.
3. J. Piella, N. G. Bastus, and V. Puntes, Z. Phys. Chem. 231 (2017) 33.
4. C. Guhrenz, A. Wolf, M. Adam, L. Sonntag, S. V. Voitekhovich, S. Kaskel, N. Gaponik, and Alexander Eychmüller, Z. Phys. Chem. 231 (2017) 51.
5. A. Freytag, M. Colombo, and N. C. Bigall, Z. Phys. Chem. 231 (2017) 63.
6. K. Boldt, Z. Phys. Chem. 231 (2017) 77.
7. P. Adel, J. Bloh, D. Hinrichs, T. Kodanek, and D. Dorfs, Z. Phys. Chem. 231 (2017) 93.
8. C. Schliehe, B. H. Juarez, M. Pelletier, S. Jander, D. Greshnykh, M. Nagel, A. Meyer, S. Forster, A. Kornowski, C. Klinke, and H. Weller, Science 329 (2010) 550.
9. S. Dogan, T. Bielewicz, V. Lebedeva, and C. Klinke, Nanoscale 7 (2015) 4875.
10. J. Lauth, F. E. S. Gorris, M. S. Khoshkhoo, T. Chassé, W. Friedrich, V. Lebedeva, A. Meyer, C. Klinke, A. Kornowski, M. Scheele, and H. Weller, Chem. Mater. 28 (2016) 1728.
11. J. Lauth, S. Kinge, and L. D. A. Siebbeles, Z. Phys. Chem. 231 (2017) 107.
12. A. Chanaewa, K. Poulsen, A. Gräfe, C. Gimmler, and E. von Hauff, Z. Phys. Chem. 231 (2017) 121.
13. C. Schedel, R. Thalwitzer, M. S. Khoshkhoo, and M. Scheele, Z. Phys. Chem. 231 (2017) 135.